\pdfoutput=1

\documentclass[pra,aps,superscriptaddress,showpacs,reprint]{revtex4-1}

\usepackage{etoolbox}
\usepackage{gensymb}

\usepackage{graphicx}
\usepackage{subfigure}
\usepackage{amsmath}
\usepackage{amssymb}
\usepackage{dsfont}

\usepackage{multirow}
\usepackage{rotating}

\usepackage{color}

\renewcommand{\Re}{\mathrm{Re}}
\renewcommand{\Im}{\mathrm{Im}}
\newcommand{\kHz}{\;\mathrm{kHz}}
\newcommand{\MHz}{\;\mathrm{MHz}}
\newcommand{\GHz}{\;\mathrm{GHz}}
\newcommand{\ns}{\;\mathrm{ns}}
\newcommand{\us}{\;\mathrm{\mu s}}
\newcommand{\panel}[1]{(\textsf{#1})}

\newcommand{\zurich}{\affiliation{Department of Physics, ETH Z\"urich, CH-8093 Z\"urich, Switzerland}}
\newcommand{\bilbao}{\affiliation{Department of Physical Chemistry, University of the Basque Country UPV/EHU, Apartado 644, E-48080 Bilbao, Spain}}

\newcommand{\beginappendix}{%
        \setcounter{table}{0}
        \renewcommand{\thetable}{A\arabic{table}}%
        \setcounter{figure}{0}
        \renewcommand{\thefigure}{A\arabic{figure}}%
     }

\begin{document}

\begin{abstract}
Systems of interacting quantum spins show a rich spectrum of quantum phases and display interesting many-body dynamics. Computing characteristics of even small systems on conventional computers poses significant challenges. A quantum simulator has the potential to outperform standard computers in calculating the evolution of complex quantum systems. Here, we perform a digital quantum simulation of the paradigmatic Heisenberg and Ising interacting spin models using a two transmon-qubit circuit quantum electrodynamics setup. We make use of the exchange interaction naturally present in the simulator to construct a digital decomposition of the model-specific evolution and extract its full dynamics. This approach is universal and efficient, employing only resources which are polynomial in the number of spins and indicates a path towards the controlled simulation of general spin dynamics in superconducting qubit platforms.
\end{abstract}

\title{Digital quantum simulation of spin models\\with circuit quantum electrodynamics}

\author{Y.~Salath\'e}
\email{ysalathe@phys.ethz.ch}
\zurich

\author{M.~Mondal}
\zurich

\author{M.~Oppliger}
\zurich

\author{J.~Heinsoo}
\zurich

\author{P.~Kurpiers}
\zurich

\author{A.~Poto\v{c}nik}
\zurich

\author{A.~Mezzacapo}
\bilbao

\author{U.~Las Heras}
\bilbao

\author{L.~Lamata}
\bilbao

\author{E.~Solano}
\bilbao
\affiliation{IKERBASQUE, Basque Foundation for Science, Maria Diaz de Haro 3, 48013 Bilbao, Spain}

\author{S.~Filipp}
\altaffiliation{Now at: IBM T. J. Watson Research Center, Yorktown Heights, NY 10598, United States}
\zurich

\author{A.~Wallraff}
\zurich

\maketitle

Quantum simulations using well controllable quantum systems to simulate the properties of another less tractable one~\cite{Feynman1982,Lloyd1996} are expected to be able to predict the properties and dynamics of diverse systems in condensed matter~\cite{Balents2010,Sachdev1999}, quantum chemistry~\cite{Lanyon2010} and high energy physics~\cite{Cirac2010,Bermudez2010}. In particular, quantum simulations are expected to provide new insights into open problems such as modeling high-Tc superconductivity~\cite{Anderson1997}, thermalization~\cite{Gogolin2011} and non-equilibrium dynamics~\cite{Polkovnikov2011}. Up to now, several prototypical quantum simulations have been proposed and realized in trapped ions~\cite{Blatt2012}, cold atoms~\cite{Bloch2012}, and quantum photonics~\cite{Aspuru-Guzik2012}. Examples include spin models~\cite{Friedenauer2008,Lanyon2011,LasHeras2014}, many-body physics~{\cite{Greiner2002}, and relativistic quantum mechanics~\cite{Gerritsma2010}. In the field of superconducting circuits quantum simulations are still in their infancy~\cite{Houck2012}. Topological properties~\cite{Roushan2014,Schroer2014} have been simulated recently, as have been fermionic models~\cite{Barends2015}.

Quantum simulators are typically classified into two main categories, namely, analog and digital ones. Analog quantum simulators are designed to display intrinsic dynamics which are equivalent to those of the simulated system. While this approach is not universal it features control of the relevant Hamiltonian parameters better than in the system to be simulated. Instead, digital quantum simulators~\cite{Lloyd1996} can reproduce the dynamics of a quantum system via a universal digital decomposition of its Hamiltonian $H=\sum_k H_k$ into efficient elementary gates realizing $H_k$. This approach is based on the Suzuki-Lie-Trotter expansion of the time evolution $U(t) = e^{-iHt} = \lim_{n\rightarrow \infty}(\prod_{k=1}^N e^{-iH_k t/n})^n$ and was recently demonstrated experimentally in a trapped-ion digital quantum simulator~\cite{Lanyon2011}.

Here we demonstrate digital quantum simulation of spin systems~\cite{LasHeras2014} in an architecture known as circuit quantum electrodynamics (QED)~\cite{Wallraff2004}.

Our experiments are carried out with two superconducting transmon qubits~\cite{Koch2007} coupled dispersively to a common mode of a coplanar waveguide resonator (see Appendix~\ref{sec:chip} for the device layout and setup diagram). We operate the circuit at $30\;\mathrm{mK}$ in a dilution refrigerator. The qubits Q1 and Q2 interact with a coplanar waveguide resonator with a fundamental resonance frequency at $7.14\GHz$ which serves both as a quantum bus~\cite{Majer2007} and for readout~\cite{Bianchetti2009}.

The natural two-qubit interaction is the XY exchange coupling~\cite{Majer2007} $H^{xy}_{1,2}=\frac{J}{2}(\sigma_1^x\sigma_2^x+\sigma_1^y\sigma_2^y)$ mediated by virtual photons in a common cavity mode, which we also refer to as the XY interaction. Here, $\sigma^{x,y}_i$ are the Pauli operators acting on qubit $i$ and $J$ denotes the effective qubit-qubit coupling strength \cite{Filipp2011a}. The XY interaction is activated by tuning the transition frequency of qubit Q1 ($5.44\GHz$) into resonance with qubit Q2 ($5.24\GHz$) for a time $\tau$ using nanosecond time scale magnetic flux bias pulses~\cite{DiCarlo2009} (see Appendix~\ref{sec:xy}). When the qubit transition frequencies are degenerate, the resonator-mediated coupling strength is spectroscopically determined to be $J=-40.4\MHz$. To make the presentation of the simulation results independent of the actual $J$, we express the interaction time $\tau$ for a given $J$ in terms of the acquired quantum phase angle $2\lvert J\rvert\tau$.
In our setup, the action of the XY gate~(Fig.~\ref{fig:1Algorithm}a) is characterized by full process tomography for a complete set of 16 initial two-qubit states and a series of 25 different interaction times $\tau$ finding process fidelities no lower than $89\,\%$ (see Appendix~\ref{sec:process}).

\begin{figure}
\includegraphics[width=\columnwidth]{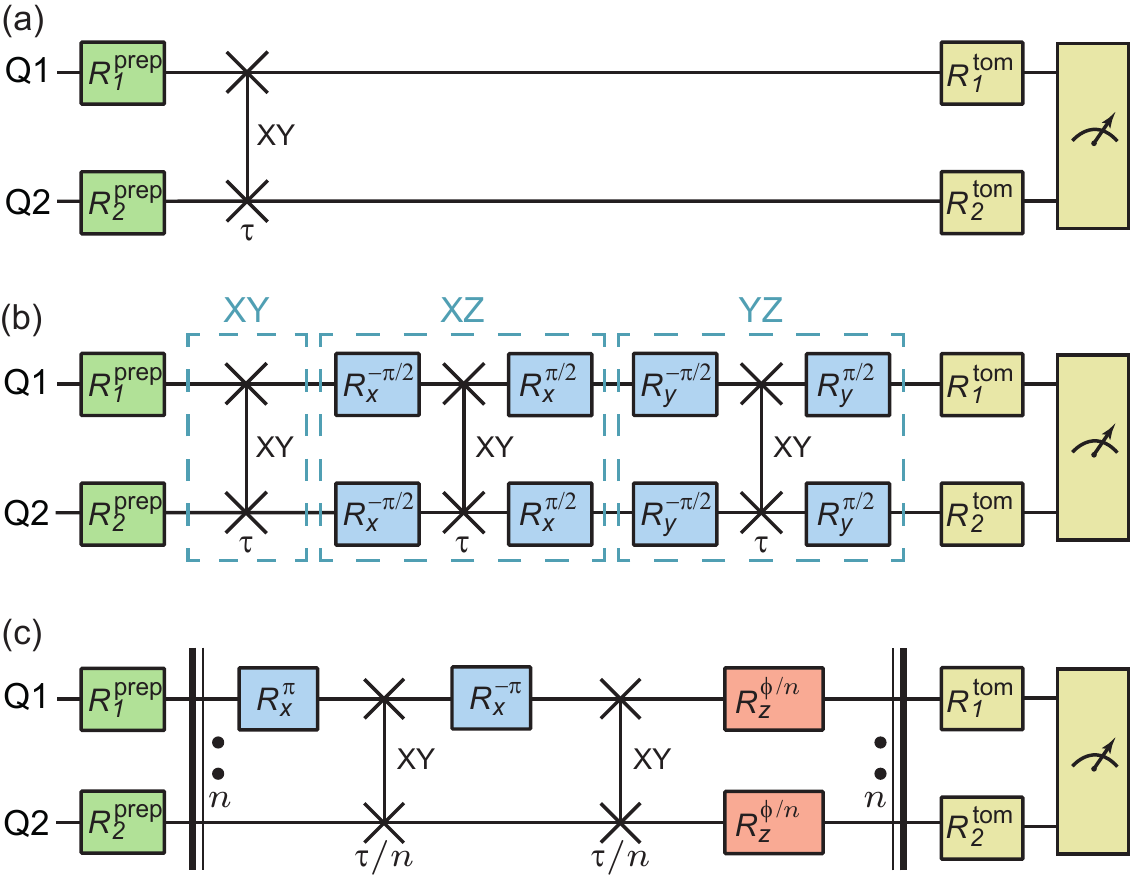}
  \caption{\panel{a}~Circuit diagram to characterize the XY exchange interaction on the qubits Q1 and Q2 symbolized by the vertical line ($\times$) which is activated for a time~$\tau$. To perform standard process tomography of this interaction, separable initial states are prepared using single-qubit rotations $R_{1,2}^\mathrm{prep}$ (green) in the beginning and the final state is characterized using single-qubit basis rotations $R_{1,2}^\mathrm{tom}$ and joint two-qubit readout (yellow).
  \panel{b}~Digital quantum simulation of the two-spin Heisenberg (XYZ) interaction for time $\tau$. The first step after state-preparation is to apply the XY gate for a time $\tau$ (dashed box labeled as XY). In the second and third steps (dashed boxes with labels XZ and YZ), XZ and YZ gates are realized using single-qubit rotations $R_{x,y}^{\pm\pi/2}$ (blue) by an angle $\pm\pi/2$ about the $x$ or $y$ axis transforming the basis in which the XY gate acts.
  \panel{c}~Protocol to decompose and simulate Ising spin dynamics in a homogeneous transverse magnetic field. 
  The circuit between the bold vertical bars with two dots is repeated $n$ times, invoking each XY and phase gates for a time $\tau/n$. See text for details. The actual pulse scheme is provided in Appendix~\ref{sec:pulses}.
  }
\label{fig:1Algorithm}
\end{figure}

In Fig.~\ref{fig:2Heisenberg}a,b we present non-stationary spin dynamics under the XY exchange interaction for a characteristic initial two-qubit state $\lvert\uparrow\rangle(\lvert\uparrow\rangle+\lvert\downarrow\rangle)/\sqrt{2}$ with spins pointing in perpendicular directions along $+\mathbf{z}$ and $+\mathbf{x}$, respectively. During the XY interaction, the state of one spin is gradually swapped to the other spin and vice versa with a phase angle of $\pi/2$.  This corresponds to the \textit{iSWAP} gate~\cite{Dewes2012}. As a consequence, the measured Bloch vectors move along the YZ and XZ planes. For a quantum phase angle of $2\lvert J\rvert\tau = \pi$ they point along the $+\mathbf{y}$ and $+\mathbf{z}$ directions respectively in good agreement with the ideal unitary time evolution indicated by dashed lines in Fig.~\ref{fig:2Heisenberg}a,b. We also find that the two-qubit entanglement characterized by the measured negativity~\cite{Vidal2002} of $0.246$ is close to the maximum expected value of $0.25$ for this initial state at a quantum phase angle of $\pi/2$. As a consequence the Bloch vectors do not remain on the surface of the Bloch sphere but rather lie within the sphere.

\begin{figure*}
\includegraphics{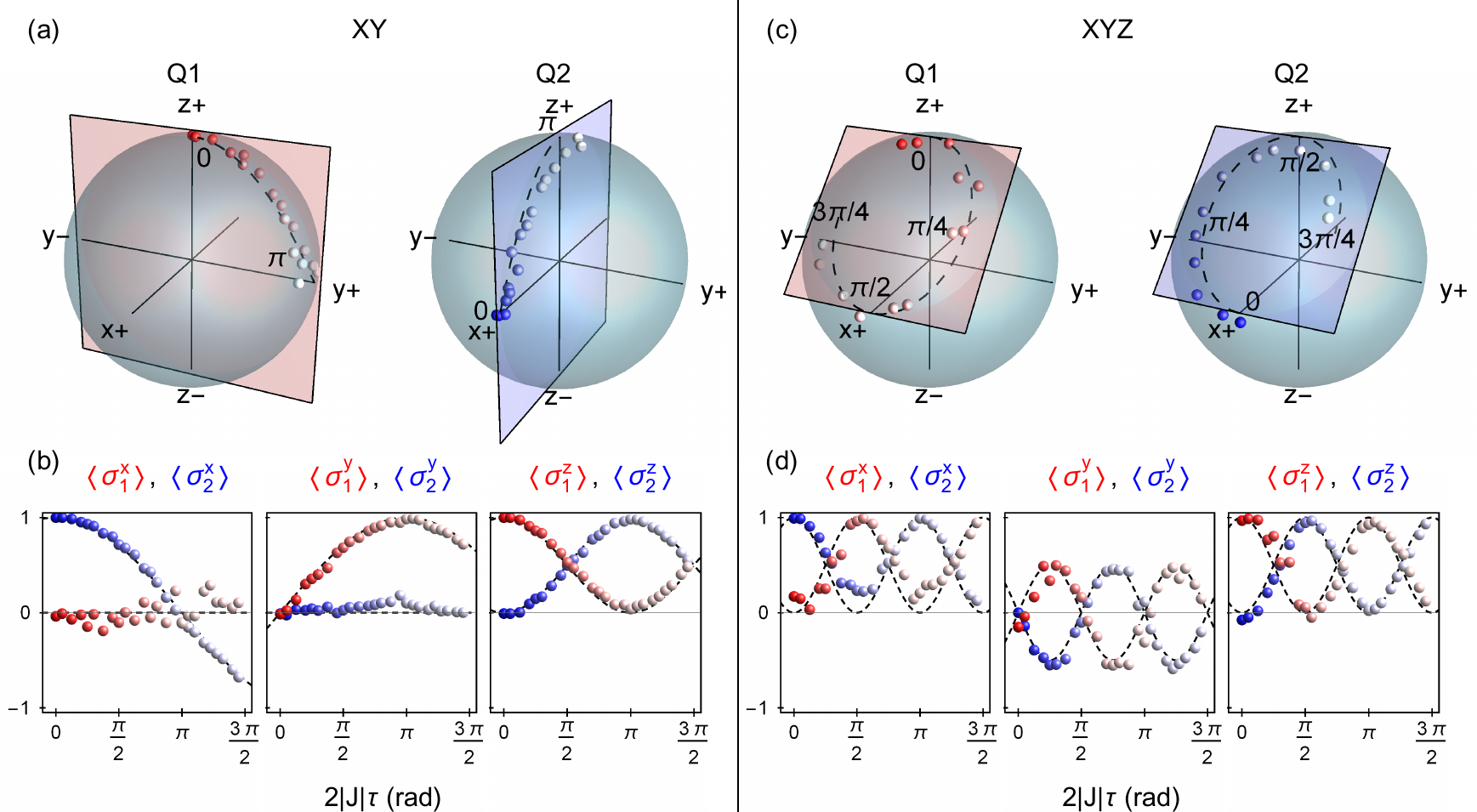}
  \caption{
  \panel{a}~Experimentally determined coordinates of the Bloch vectors during exchange (XY) interaction represented by small red~(Q1) and blue~(Q2) points are compared to the ideal paths shown as dashed lines in the XY model. The ideal paths are in the YZ and XZ planes shown as blue and red planes intersecting the Bloch sphere. The time evolution is indicated by the saturation of the colors
  as the quantum phase angle~$2\lvert J\rvert\tau$ advances from $0$ (saturated) to $\pi$ (unsaturated). \panel{b}~Measured expectation values of the Pauli operators~$\sigma^{x,y,z}_{1,2}$ for the qubits Q1 (red points) and Q2 (blue points), respectively, for the XY interaction as a function of the quantum phase angle $2\lvert J\rvert\tau$ along with the ideal evolution (dashed line). \panel{c}~Evolution of the Bloch vector for the quantum simulation of the isotropic Heisenberg interaction vs. quantum phase angles from $0$ to $3\pi/4$. The path of the Bloch vectors of the qubits Q1 and Q2 spans the plane indicated by the rectangular sheets intersecting the Bloch spheres. \panel{d},~As in panel \textbf{b} for the Heisenberg interaction.}

\label{fig:2Heisenberg}
\end{figure*}

The anisotropic Heisenberg model describes spins interacting in three spatial dimensions
 \begin{equation}\label{eq:heisenberg}
H_{xyz}=\sum_{(i,j)} (J_x\sigma_i^x\sigma_j^x+J_y\sigma_i^y\sigma_j^y+J_z\sigma_i^z\sigma_j^z),
\end{equation}
where the sum is taken over pairs of neighbouring spins $i$ and $j$. $J_x$, $J_y$ and $J_z$ are the couplings of the spins along the $x$, $y$ and $z$ coordinates, respectively.
Since it does not occur naturally in circuit QED we decompose the Heisenberg interaction into a sequence of XY and single-qubit gates, as shown in Fig.~\ref{fig:1Algorithm}b. We combine three successive effective XY, XZ and YZ gates derived from the XY gate by basis transformations~\cite{LasHeras2014} to realize the isotropic Heisenberg model with $J_x=J_y=J_z=J$ versus interaction time $\tau$. Since the XY, XZ and YZ operators commute for two spins the Trotter formula is exact after a single step.

To compare the Heisenberg (XYZ) interaction with the XY exchange interaction we have prepared the same initial state as presented in Fig.~\ref{fig:2Heisenberg}a,b. The isotropic Heisenberg interaction described by the scalar product between two vectorial spin $1/2$ operators preserves the angle between the two spins. As a result, the initially perpendicular Bloch vectors of qubits Q1 and Q2 remain perpendicular during the interaction~(Fig.~\ref{fig:2Heisenberg}c) and rotate clockwise along an elliptical path that spans a plane perpendicular to the diagonal at half angle between the two Bloch vectors~(Fig.~\ref{fig:2Heisenberg}c).

In accordance with theory, the XYZ interaction leads to a full \textit{SWAP} operation for a quantum phase angle of $2\lvert J\rvert\tau=\pi/2$ where the Bloch vectors point along the $+\mathbf{x}$ and $+\mathbf{z}$ directions.
For the given initial state, we observed a maximum negativity of $0.210$ close to the expected value of $0.25$ for the Heisenberg interaction at a quantum phase angle of $2\lvert J\rvert\tau = \pi/4$. As for the XY interaction we have characterized the Heisenberg interaction with standard process tomography finding fidelities above $82\,\%$ for all quantum phase angles $2\lvert J\rvert\tau$.

Next, we consider the quantum simulation of the Ising model with a transverse homogeneous magnetic field
\begin{equation}\label{eq:ising}
H_{I} = J\sum_{(i,j)} \sigma_i^x\sigma_j^x +  \frac{B}{2} \sum_i \sigma_i^z,
\end{equation}
where the magnetic field $B$ pointing along the $z$ axis is perpendicular to the interaction given by $J\sigma^x_i\sigma^x_j$. Since the two-spin evolution~(Fig.~\ref{fig:1Algorithm}c) is decomposed into two-qubit XY and single-qubit Z gates which do not commute, the transverse field Ising dynamics is only recovered using the Trotter expansion in the limit of a large number of steps~$n$ for an interaction time of $\tau/n$ in each step. To realize the Ising interaction term using the exchange interaction, the XY gate is applied twice for a time $\tau/n$, once enclosed by a pair of $\pi$ pulses on qubit Q1. This leads to a change of sign of the $\sigma^y_1\sigma^y_2$ term which thus gets canceled when added to the bare XY gate. The external magnetic field part of the Hamiltonian is realized as single-qubit phase gates~$R_z^{\phi}$ which rotate the Bloch vector about the $z$ axis by an angle~$\phi=B\tau/n$ per Trotter step. These gates are realized by detuning the respective qubit by an amount $\delta$ from its idle frequency corresponding to an effective $B$-field strength of $B=2\pi\delta$.

We experimentally simulate the non-stationary dynamics of two spins in this model for the initial state $\lvert \uparrow\rangle(\lvert\uparrow\rangle-i\lvert\downarrow\rangle)/\sqrt{2}$ which is well-suited to assess the simulation performance. In Fig.~\ref{fig:3Ising}a expectation values for the digital simulation of the $\sigma^z_{1,2}$-components of the two spins are shown, as well as the two-point correlation function $\langle\sigma^x_1\sigma^x_2\rangle$. The $\sigma^z_{1,2}$-components of the spins represented by the red and blue datasets in Fig.~\ref{fig:3Ising}a, respectively, oscillate with a dominant frequency component of $2J$ due to the presence of the interaction term $\propto \sigma^x_1\sigma^x_2$. Likewise, the XX correlation $\langle\sigma^x_1\sigma^x_2\rangle$ represented by the yellow dataset in Fig.~\ref{fig:3Ising}a is non-stationary and oscillates at rate $2\sqrt{B^2+J^2}=2\sqrt{10}J\approx6.3J$ due to the presence of a magnetic field of strength $B=3J$. The evolution of the measured final state shows agreement with a theoretical model (solid lines in Fig.~\ref{fig:3Ising}a) which takes into account dissipation and decoherence with deviations being dominated by systematic gate errors (see Appendix~\ref{sec:error}).

In Fig.~\ref{fig:3Ising}b the fidelity of the simulated state is compared to the expected state at characteristic quantum phase angles both for the experimental realization (colored bars) and the ideal Trotter approximation (wire frames) after the $n$th step. In an ideal digital quantum simulator the theoretical fidelity (wire frame) converges for an increasing number of steps~$n$~(Fig.~\ref{fig:3Ising}b). The experimental fidelity, however, reaches a maximum for a finite number of steps (Fig.~\ref{fig:3Ising}b) after which it starts to decrease due to gate errors and decoherence~\cite{LasHeras2014}. As expected, the Trotter approximation converges faster for smaller quantum phase angles $2\lvert J\rvert\tau$. For $2\lvert J\rvert\tau = \pi/4$ the peak experimental fidelity~(Fig.~\ref{fig:3Ising}b) of $98.3\,\%$ is already observed for $n=1$, whereas for $2\lvert J\rvert\tau = 3\pi/2$ the optimum of $80.7\,\%$ is observed for $n=5$.

\begin{figure*}
\includegraphics{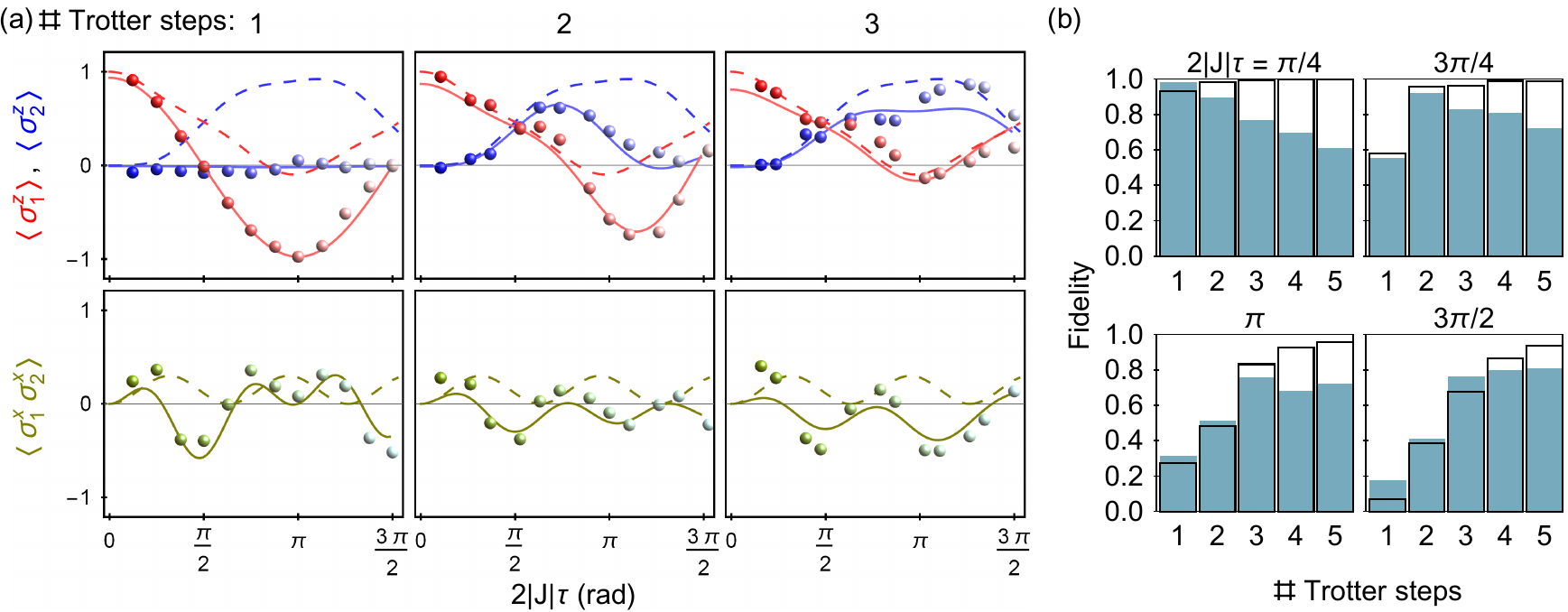}
  \caption{
  \panel{a}~Digital quantum simulation of the Ising model with transverse homogeneous magnetic field using 1 to 3 Trotter steps. Shown are the $z$-components $\langle\sigma^z_1\rangle$ of qubit~Q1 (red) and $\langle\sigma^z_2\rangle$ of qubit~Q2 (blue) and the two-point correlation function in the $x$-direction $\langle\sigma^x_1\sigma^x_2\rangle$ (yellow points) of the spins as a function of the quantum phase angle $2\lvert J\rvert\tau$ for the initial state $\lvert\uparrow\rangle(\lvert\uparrow\rangle-i\lvert\downarrow\rangle)/\sqrt{2}$ and a magnetic field strength $B=3J$. Theoretically expected results take systematic phase offsets and finite coherence of the qubits into account (solid curves). The ideal dynamics are obtained from the time-dependent Schr\"odinger equation for the Ising Hamiltonian (dashed lines). \panel{b}~Fidelity with respect to the exactly solved Ising model for displayed quantum phase angles of the final state after ideal unitary evolution in the simulation protocol for~$n$ Trotter steps (wire frames) and experimentally obtained final state (colored bars).}

\label{fig:3Ising}
\end{figure*}

In future experiments, transmission line resonators may provide a means to design multi-qubit devices with non-local qubit-qubit couplings that directly reflect the lattice topology of spin systems such as frustrated magnets. Moreover, the incorporation of cavity modes as explicit degrees of freedom in the simulated models~\cite{Mezzacapo2014a}, following an analog-digital approach, and the integration of optimal control concepts, will be instrumental to scale the system to larger Hilbert-space dimensions. With this, the circuit QED architecture offers considerable potential for surpassing the limitations of classical simulations, which can be facilitated by using efficient digital decompositions of spin Hamiltonians as pursued in this work.

\textbf{Acknowledgments} The authors would like to thank Abdufarrukh Abdumalikov and Marek Pechal for helpful discussions. Furthermore we owe gratitude to Lars Steffen, Arkady Fedorov, Christopher Eichler, Mathias Baur, Jonas Mlynek who contributed to our experimental setup. We would also like to thank Tim Menke and Andreas Landig for contributions to the calibration software used in the present experiment.

We acknowledge financial support from Eidgen\"ossische Technische Hochschule Zurich (ETH Zurich), the Swiss National Science Foundation's National Centre of Competence in Research `Quantum Science \& Technology', the Basque Government IT472-10, Spanish MINECO FIS2012-36673-C03-02, Ram\'on y Cajal Grant RYC-2012-11391, UPV/EHU Project No. EHUA14/04, PROMISCE and SCALEQIT European projects.

\beginappendix
\appendix

\section{Chip architecture and measurement setup}\label{sec:chip}
The present experiment was performed using two superconducting transmon~\cite{Koch2007} qubits Q1 and Q2 and one coplanar waveguide resonator R1 on a microchip (Fig.~\ref{fig:S1DeviceChip}). The resonator R1 has a fundamental resonance frequency of $\nu_r = 7.14\GHz$. From spectroscopic measurements we have determined the maximum transition frequencies $\nu_\mathrm{max}= \{5.55,\,5.53\}\GHz$ and charging energies $E_C/h \approx\{260,\,260\}\MHz$ of the qubits Q1 and Q2, respectively, where $h$ is the Planck constant. The qubits Q1 and Q2 are coupled to resonator R1 with coupling strengths $g/2\pi \approx\{120,\,120\}\MHz$. For this experiment the qubit transition frequencies in their idle state were offset to $\nu= \{5.440,\,5.240\}\GHz$ by applying a constant magnetic flux threading their SQUID loops with miniature superconducting coils mounted underneath the chip. At these idle frequencies, the measured energy relaxation and coherence times were $T_1 = \{7.1,\,6.7\}\us$ and $T_2 = \{5.4,\,4.9\}\us$, respectively. The transition frequencies of the qubits Q3 and Q4 were tuned to $4.5\GHz $ and $6.1\GHz $ such that they do not interact with Q1 and Q2 during the experiment.

\begin{figure*}[!ht]
\centerline{\includegraphics[width=16cm]{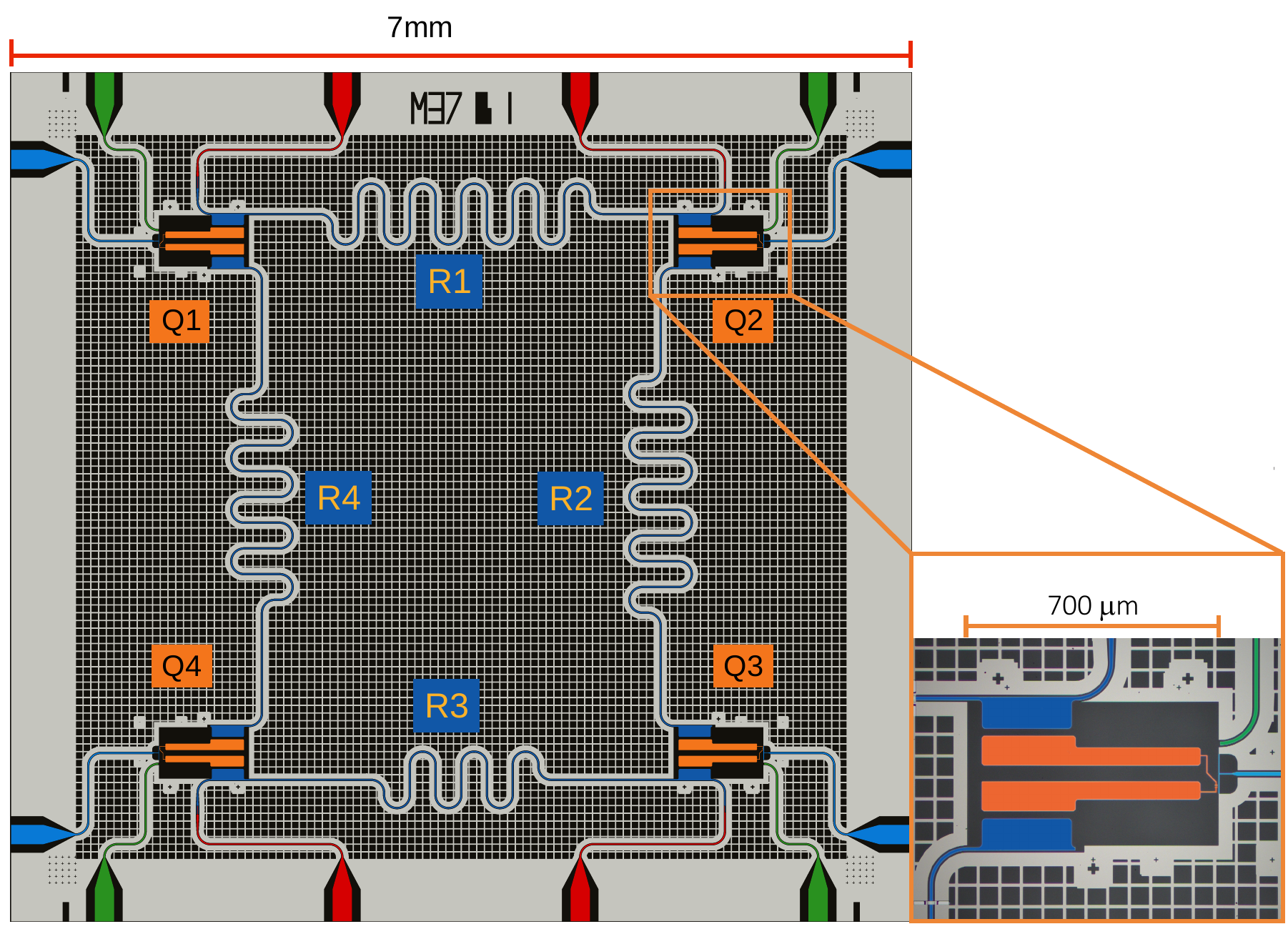}}
\caption{Chip design and false colored optical image of a superconducting qubit (inset). The chip comprises four superconducting qubits Q1-4 (orange) made of aluminium and four niobium coplanar waveguide resonators R1-4 (deep blue) coupled to input and output ports (red). The qubits have individual microwave drive lines (green) and flux bias lines (blue).}
\label{fig:S1DeviceChip}
\end{figure*}

A schematic diagram of the measurement setup is shown in Fig.~\ref{fig:S2DeviceSchematic}a. To realize two-qubit XY gates and single-qubit phase gates (Z), controlled voltage pulses generated by an arbitrary waveform generator~(AWG) are used to tune the flux threading the SQUID loop of each qubit individually using flux bias lines~\cite{DiCarlo2009}.
The single-qubit microwave pulses (X,Y) are generated using sideband modulation of an up-conversion in-phase quadrature~(IQ) mixer (Fig.~\ref{fig:S2DeviceSchematic}b) driven by a local oscillator~(LO) and modulated by an arbitrary waveform generator~(AWG). The same up-conversion LO is used for the microwave pulses on both qubits to minimize the phase error introduced by phase drifts of microwave generators. We have used a quantum-limited parametric amplifier~(PA) to amplify readout pulses at the output of R1 (Fig.~\ref{fig:S2DeviceSchematic}c). Here the Josephson junction based amplifier  in form of a Josephson parametric dimer~(JPD)~\cite{Eichler2014a} is pumped by a strong pump drive through a directional coupler (D). To cancel the pump leakage, a phase~($\phi$) and amplitude~(A) controlled microwave cancelation tone is coupled to the other port of the directional coupler~(D). Three circulators~(C1-3) were used to isolate the sample from the pump tone. A circulator (C4) at base temperature followed by a cavity band-pass filter (BP) and another circulator (C5) at the still stage were used to isolate the sample and JPD from higher-temperature noise. The transmitted signal is further amplified by a high electron mobility transistor~(HEMT) at the $4.2\;\mathrm{K}$ stage and a chain of ultra-low-noise~(ULN) and low-noise~(LN) amplifiers at room temperature as shown in~Fig.~\ref{fig:S2DeviceSchematic}d. The amplified readout pulse is down-converted to an intermediate frequency~(IF) of $25\MHz$ using an IQ mixer~(Fig.~\ref{fig:S2DeviceSchematic}e) and digitally processed by field-programmable gate array~(FPGA) logic for real-time data analysis.

\begin{figure*}[!ht]
\centerline{\includegraphics[width=16cm]{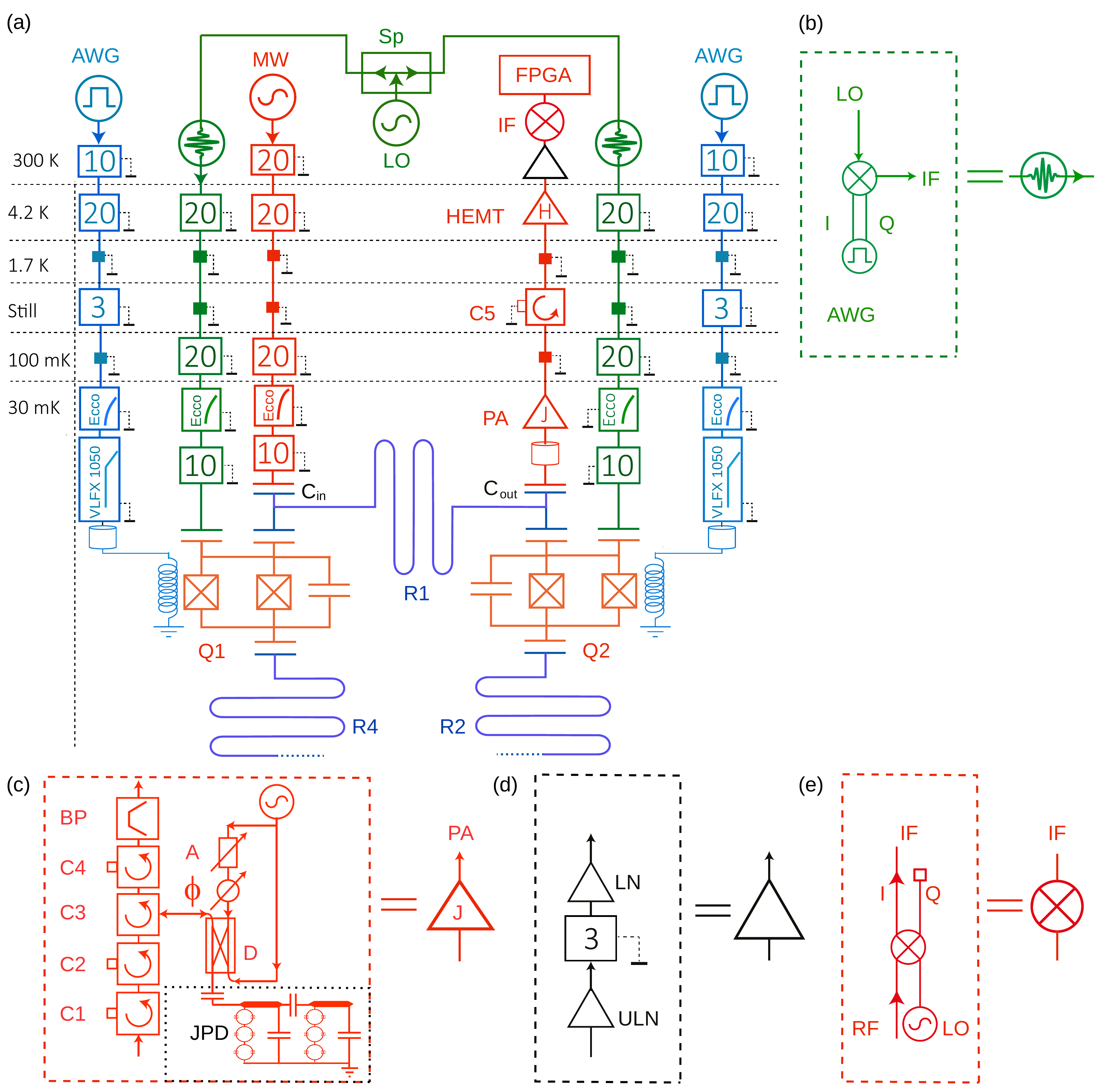}}
\caption{\panel{a}~Schematic of the experimental setup with complete wiring of electronic components inside and outside of the dilution refrigerator
with the same color code as in Fig.~\ref{fig:S1DeviceChip}. \panel{b}~Up conversion circuit for generating controlled microwave pulses. \panel{c}~Quantum limited parametric amplifier circuit to amplify readout pulses at base temperature. \panel{d}~Amplifiers used at room temperature just before down conversion of the signal. \panel{e}~Down conversion circuit (See text for details).}
\label{fig:S2DeviceSchematic}
\end{figure*}

\section{Implementation of the XY gate}\label{sec:xy}
The interaction between two qubits with degenerate transition frequencies dispersively coupled to the same CPW resonator is described by the exchange coupling~\cite{Blais2004} $J(\sigma^+_1\sigma^-_2 + \sigma^-_1\sigma^+_2)$ which can also be written in terms of Pauli operators as $\frac{J}{2}(\sigma_1^x\sigma_2^x+\sigma_1^y\sigma_2^y)$. We activate this interaction by tuning the transition frequency of qubit Q1 into resonance with qubit Q2 with a flux pulse (Fig.~\ref{fig:S3XYgate}) for an interaction time $\tau$ which we varied from $0$ to $60\ns$. At the frequency of qubit Q2, we obtain a coupling strength $J=-40.4\MHz$ from a fit to the spectroscopically measured avoided crossing. To compensate overshoots of the flux pulse due to the limited bandwidth of the flux line channel, we use an inverted linear filter based on room-temperature response measurements of the flux line channel and in-situ Ramsey measurements of the residual detuning of qubit Q1 in the time interval from $0$ to $2\us$ after the flux pulse.

\begin{figure}[!ht]
\centerline{\includegraphics[width=8.9cm]{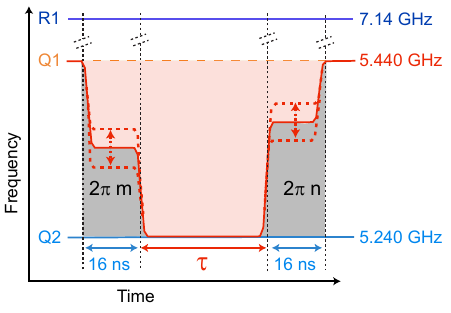}}
\caption{Implementation of the XY gate. The transition frequency of qubit Q1 (red) is tuned into resonance with qubit Q2 (blue) for an interaction time $\tau$ using a fast flux pulse. Before and after the flux pulse, a $16\ns$ long buffer is added at an intermediate level to cancel the dynamic phase accumulated by qubit Q1 relative to Q2 (grey area) during the evolution (see text).}
\label{fig:S3XYgate}
\end{figure}

Since the outcome of the XY gate depends strongly on the relative phase of the two-qubit input state, we have used the same LO signal for the upconversion of the single-qubit pulses acting on both qubits Q1 and Q2 (green lines in Fig.~\ref{fig:S2DeviceSchematic}a). Then the initial relative phase between the qubits is defined solely by the pulse sequence generated by the AWG and the cable lengths. In addition, we choose the shape of the flux pulse that realizes the XY gate such that the dynamic phase acquired by qubit Q1 during the idle time and the rising edge of the flux pulse cancels any unwanted relative phase offset of the initial state. We satisfy this condition by tuning the frequency of Q1 to an intermediate level (buffer) for a fixed time of $16\ns$ before and after the XY gate (Fig.~\ref{fig:S3XYgate}). A suitable buffer level is found by performing Ramsey-type experiments with a single XY gate while sweeping the buffer amplitudes. This calibration procedure is carried out for each interaction length of the XY gate.
The second buffer at the falling edge of the flux pulse is used to ascertain that the relative phase between the qubits after tuning qubit Q1 back to its original position is the same as the initial relative phase.

\section{Pulse scheme}\label{sec:pulses}

The quantum protocols for the digital quantum simulation of Heisenberg (Fig.~\ref{fig:S4pulseSchemes}\textbf{a}) and Ising spin (Fig.~\ref{fig:S4pulseSchemes}\textbf{b}) models were realized by sequences of microwave and flux pulses applied on qubit Q1~(red curves in~Fig.~\ref{fig:S4pulseSchemes}) and qubit Q2~(blue curves in~Fig.~\ref{fig:S4pulseSchemes}). The single-qubit rotations were implemented by $24\ns$ long Gaussian-shaped resonant DRAG~\cite{Motzoi2009,Gambetta2011a} microwave pulses and the XY gates were implemented using fast flux pulses. To avoid the effect of residual transient response of the flux pulse we have added a $40 \ns + \Delta\tau$ waiting time after each flux pulse, with $\Delta\tau$ being an adjustable idle time. We have chosen $\Delta\tau$ such that the time difference between two applications of the XY interaction is commensurate with the relative phase oscillation period of $5\ns$, equal to the inverse frequency detuning $1/200\MHz$. With these measures we ensure that the gate can be used in a modular fashion, i.e. that a single calibration of the gate suffices for all gate realizations within the algorithm. The single-qubit phase gates were implemented by detuning the idle frequencies of each qubit with a square flux pulse. In the idle state, we observe a state-dependent qubit transition frequency shift of $940\kHz$ due to the residual $\sigma^z_1\sigma^z_2$ interaction. To decouple this undesired effect we have used a standard refocusing technique~\cite{Vandersypen2004} implemented by two consecutive $\pi$ pulses on qubit Q2 (magenta boxes in~Fig.~\ref{fig:S4pulseSchemes}). In the end of each pulse sequence we perform dispersive joint two-qubit state-tomography~\cite{Filipp2009b} by single-qubit basis transformations followed by a pulsed microwave transmission measurement through resonator R1.

\begin{figure*}[!ht]
\centerline{\includegraphics{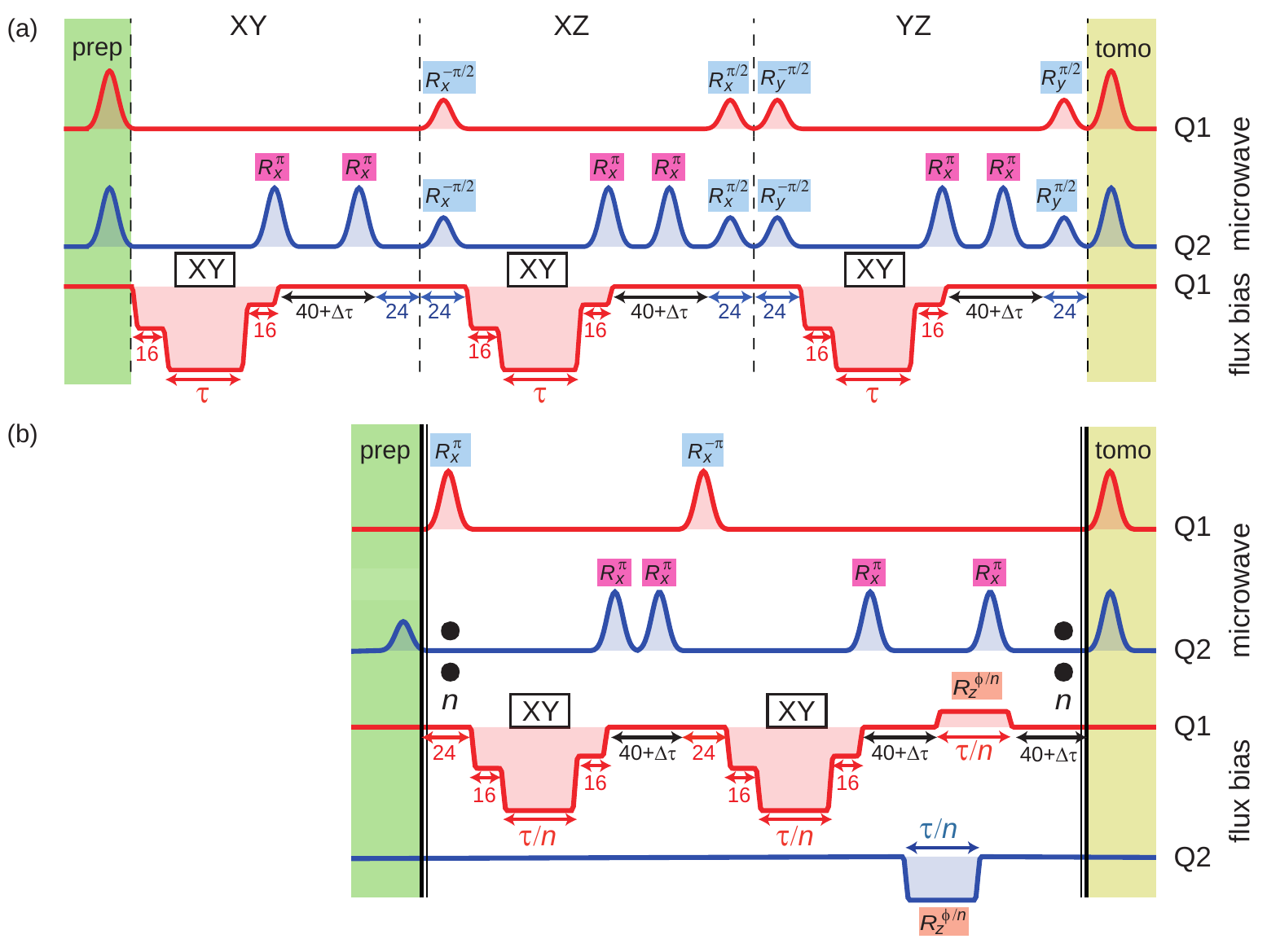}}
\caption{Pulse sequences that are applied on qubit Q1 (red) and qubit Q2 (blue) to implement the Heisenberg \panel{a} and Ising spin \panel{b} models. The Gaussian-shaped DRAG microwave pulses are applied to the charge lines of the respective qubits to implement single-qubit rotations $R_{x,y}^\phi$ about the $x$ or $y$ axis of the Bloch vector by an angle $\phi$. Each sequence starts with the preparation of an initial state (green boxes) and ends with microwave pulses for basis rotations to perform state-tomography (yellow boxes). The microwave pulses marked with magenta boxes are used for refocussing. The black vertical bars with the two dots in panel \panel{b} indicate that the enclosed pulse sequence is repeated $n$ times. The XY gates are realized by applying flux pulses to the flux line of qubit Q1 for a time $\tau/n$. The phase gates $R_z^{\phi/n}$ are implemented by detuning the transition frequency of each qubit from their idle frequencies applying flux pulses for a time $\tau/n$. The numbers stated below the pulses on qubit Q1 represent timescales in ns.}

\label{fig:S4pulseSchemes}
\end{figure*}

\section{Process tomography}\label{sec:process}

We perform standard two-qubit process tomography~\cite{Poyatos1997,Chuang1997} of the XY gate and of the simulated isotropic Heisenberg (XYZ) model for a varying interaction time $\tau$. Fig.~\ref{fig:S5processXY} shows the process $\chi$ matrices characterizing the XY gate for a quantum phase angle $\pi/2$~(Fig.~\ref{fig:S5processXY}a) and $\pi$~(Fig.~\ref{fig:S5processXY}b) corresponding to a $\sqrt{\textit{iSWAP}}$ gate~\cite{Kofman2009,Dewes2012} and an \textit{iSWAP} gate~\cite{Schuch2003,Neeley2010a} with process fidelities of $97.8\,\%$ and $95.3\,\%$, respectively. Heisenberg interaction with a quantum phase angle $\pi/2$ leads to a \textit{SWAP} gate~(Fig.~\ref{fig:S6processXYZ}a) with a process fidelity of $86.1\,\%$. While the SWAP gate belongs to the two-qubit Clifford group, there is no natural interaction in standard circuit QED architecture to directly implement the SWAP gate~\cite{Wu2012d,Corcoles2013}.
For a phase angle $\pi$, the Heisenberg interaction is an identity gate~(Fig.~\ref{fig:S6processXYZ}b) with a process fidelity of $83.6\,\%$.

\begin{figure*}
\centerline{\includegraphics[width=15.5cm]{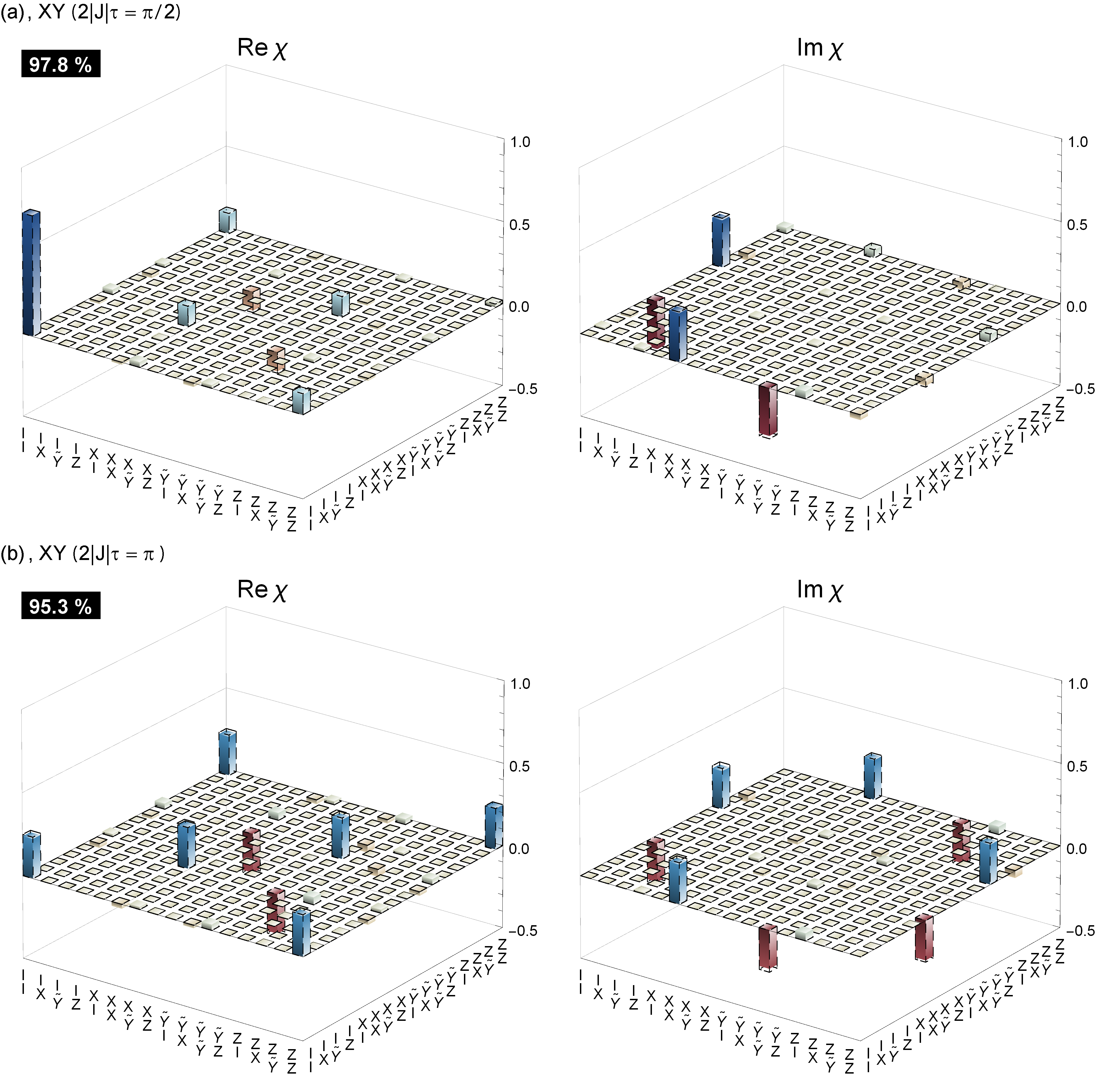}}
\caption{\panel{a}~Measured real and imaginary part of the XY process $\chi$ matrix ($\Re\ \chi$, $\Im\ \chi$), in the basis $\{I=\textit{identity},\,X=\sigma_x,\,\tilde{Y}=-i\sigma_y,\,Z=\sigma_z\}$, describing the mapping from any initial state to the final state for a quantum phase angle of $2\lvert J\rvert\tau = \pi/2$. The dashed wire frames represent the theoretically optimal matrix elements and the colored bars represent measured positive (blue) and negative (red) matrix elements. The fidelity of the experimentally observed process with respect to the ideal process is indicated in the black boxes. \panel{b}~As in \panel{a} for a phase angle $\pi$.
}
\label{fig:S5processXY}
\end{figure*}

\begin{figure*}
\centerline{\includegraphics[width=15.5cm]{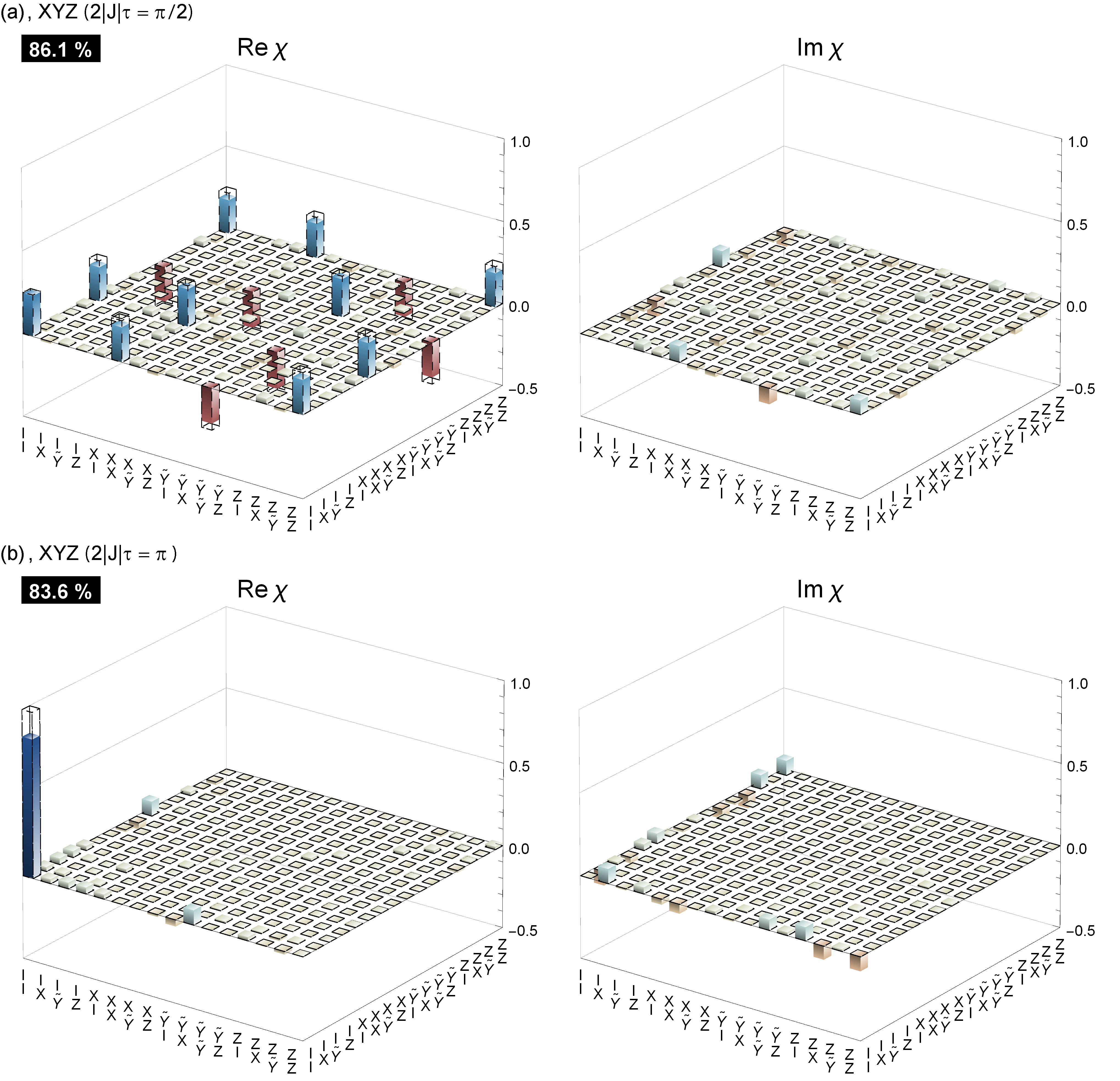}}
\caption{\panel{a}~Measured real and imaginary part of the Heisenberg (XYZ) process $\chi$ matrix ($\Re\ \chi$, $\Im\ \chi$), in the basis $\{I=\textit{identity},\,X=\sigma_x,\,\tilde{Y}=-i\sigma_y,\,Z=\sigma_z\}$, describing the mapping from any initial state to the final state for a quantum phase angle of $2\lvert J\rvert\tau = \pi/2$. The dashed wire frames represent the theoretically optimal matrix elements and the colored bars represent measured positive (blue) and negative (red) matrix elements. The fidelity of the experimentally observed process with respect to the ideal process is indicated in the black boxes. \panel{b}~As in \panel{a} for a phase angle $\pi$.
}
\label{fig:S6processXYZ}
\end{figure*}

\section{Error contributions}\label{sec:error}

The single-qubit gate fidelities measured by randomized benchmarking~\cite{Knill2008,Chow2009,Kelly2014} amount to~$99.7\,\%$. The dominant contribution to the loss in fidelity originates from the two-qubit XY gates for which a process fidelity $\mathcal{F}_\mathrm{p,XY} = 95.7\,\%$ is obtained from process tomography averaging over all quantum phase angles. This indicates that the errors in the implementation of the XY gate limit the fidelity of the final state of the quantum simulation. To confirm this, we calculate the expected process fidelity for the Heisenberg and Ising protocol from the observed XY gate fidelity by assuming independent gate errors in all three steps. For the Heisenberg (XYZ) model simulation neglecting the small single-qubit gate errors, we expect a mean process fidelity $\mathcal{F}_\mathrm{p,XYZ} \approx 1-3(1-\mathcal{F}_\mathrm{p,XY})=87.1\,\%$, which is close to the observed value of $86.3\,\%$. For the Ising model simulation we expect a process fidelity of $\mathcal{F}_\mathrm{p,Ising} \approx 1-2n(1-\mathcal{F}_\mathrm{p,XY})$. From the relation $\mathcal{F}_\mathrm{s}=(d \mathcal{F}_\mathrm{p}+1)/(d+1)$ between state ($\mathcal{F}_s$) and process fidelity ($\mathcal{F}_p$), we obtain the expected mean state fidelities of $\{93.1, 86.2, 79.4, 72.5, 65.6\}\,\%$ for $n=1$ to $5$ Trotter steps which compare well to the measured state fidelites $\{91.7, 88.3, 82.2, 73.0, 60.7\}\,\%$.

To estimate the dominant source of systematic errors, we consider a model which includes relaxation ($T_1$) and dephasing ($T_2$) and state-dependent phase errors described by an effective $\tilde J_z\sigma^z_1\sigma^z_2$ term with interaction strength $\tilde J_z$. In addition, we include an extra offset in the single-qubit phase gate acting on qubit Q2 from cross talk of the flux pulses acting on qubit Q1 in each Trotter step.  By fitting the final state predicted by this model to the observed states, we estimate an unwanted interaction angle $\tilde J_z\tau_z$ of approximately $2.3^{\circ}$ and a constant phase offset of $4.6^{\circ}$.

\end{document}